\documentclass[a4paper,11pt]{article}

\usepackage[dvips]{graphicx}

\author{
        Oriol Vendrell,
        Fabien Gatti and
        Hans--Dieter Meyer*
       }

\title{{\bf Dynamics and Infrared Spectroscopy of the Protonated Water Dimer}**}

\date{}

\begin{document}

\maketitle

\noindent Catch phrase: {\bf Hydrated Proton}

   \begin{itemize}
   \item[[*]] Dr. O. Vendrell, Prof. Dr. H.-D. Meyer\\
   Theoretische Chemie,\\
   Physikalisch-Chemisches Institut,\\
   Universitaet Heidelberg,\\
   Im Neuenheimer Feld 229,\\
   69120 Heidelberg, Germany\\
   fax: (+49)-6221-54-5221\\
   e-mail: hans-dieter.meyer@pci.uni-heidelberg.de\\
   homepage:\\ http://www.pci.uni-heidelberg.de/tc/usr/dieter\\
   \item[] Dr. F. Gatti,\\
   CTMM, Institut Charles Gerhardt, UMR 5253,\\
   CC 014, Universit\'e Montpellier II,\\
   F - 34095 Montpellier, Cedex 05, France
   \end{itemize}

   \begin{itemize}
   \item[[**]]
    The authors thank Prof. J. Bowman for providing the potential-energy routine, 
    D. Lauvergnat for performing the TNUM calculations
    and the Scientific Supercomputing Center Karlsruhe for generously providing computer time. 
    O. V. is grateful to the Alexander von Humboldt Foundation for financial support.
   \end{itemize}

  \noindent Accurate infrared (IR) spectroscopy of protonated
  water clusters prepared in the gas phase has become possible in recent 
  years,\cite{asm03:1375,fri04:9008,hea04:11523,jia00:1398,hea05:1765,ham05:244301} opening the
  door to a deeper understanding of the properties of aqueous systems and the
  hydrated proton, which are of main interest in central areas of chemistry and biology.
  Several computational studies have appeared in parallel, providing a necessary
  theoretical basis for the assignment and understanding of the different
  spectral features.\cite{ven01:240,dai03:6571,hua05:044308,ham05:244301,sau:kal}
  It has been recently demonstrated that the {\mbox{H$_5$O$_2^+$}} motif,
  also referred to as Zundel cation,
  plays an important role
  in protonated water clusters of 6 or more water molecules
  and as a limiting structure, together with the Eigen cation (\mbox{H$_9$O$_4^+$}),
  of the hydrated proton in bulk water.\cite{jia00:1398,hea05:1765,mar99:601}
  The importance of the hydrated proton and the ammount of work devoted to the problem
  contrast with the fact that the smallest system in which a proton is shared between
  water molecules, {\mbox{H$_5$O$_2^+$}}, is not yet completely understood,
  and an explanation of the most important spectral signatures and the associated
  dynamics of the cluster is lacking. 
  
  In this letter we report the simulation of the IR linear absorption spectrum 
  of the {\mbox{H$_5$O$_2^+$}} cation in the range between 0 to 4000 {cm$^{-1}$} by state-of-the-art 
  quantum-dynamical methods, we discuss the spectral signatures in terms of the underlying couplings
  and dynamics of the different degrees of freedom and compare our results
  to recent, accurate experiments on this system.
  For the first time the doublet-peak feature at around 1000 {cm$^{-1}$} is fully
  reproduced, analyzed and assigned.
  The doublet is found to arise from the coupling between the proton-transfer
  mode and low frequency, large-amplitude displacements of both water molecules.
  Predictions are also made for the lowest frequency part of the spectrum, which
  has not yet experimentally been accessed.
  Several important features of the system are analyzed for the first time, 
  namely the degeneracy of some of the vibrational levels and the extreme anharmonicity 
  of the wagging motions and relative internal rotation of the two water molecules.
  In doing so, we do not resort to any low-dimensional model of the system, but
  we treat it in its full dimensionality, i. e.,
  $3N-6=15$ active internal coordinates (15D).
  The use of full dimensionality is found crucial in the reproduction of the complete
  absorption spectrum and dynamics.
  Our study provides a picture of the {\mbox{H$_5$O$_2^+$}} system, extendable to larger
  aggregates, in which the clusters have to be viewed as highly anharmonic, 
  flexible, multi-minima, coupled systems.
  From a methods perspective,
  we show that a full quantum-dynamical description of such a
  complex molecular system can still be achieved, providing 
  explicative and predictive power and a
  very good agreement to available experimental data.
  In this respect, the reported simulations set a new state of the art
  in {\em quantum dynamically} describing an anharmonic, highly coupled 
  molecular system of the size of the {\mbox{H$_5$O$_2^+$}} cation. 
  To account for the interatomic potential and the interaction with the radiation
  we make use of the potential energy surface (PES) 
  and dipole-moment surfaces recently developed by Bowman and collaborators,\cite{hua05:044308}
  which constitute the most accurate {\em ab initio} surfaces
  available to date for this system.
 
  The IR predissociation spectrum of the {\mbox{H$_5$O$_2^+$}} cation has been recently
  measured in argon-solvate \cite{hea04:11523} and neon- and
  argon-solvate \cite{ham05:244301} conditions.
  It is expected that the photodissociation spectrum of the {\mbox{H$_5$O$_2^+$}}$\cdot$Ne$_1$ complex is close
  to the linear absorption spectrum of the bare cation.\cite{ham05:244301}
  This spectrum features a doublet structure in the region of 1000 {cm$^{-1}$} 
  made of two well-defined absorptions at 
  928 {cm$^{-1}$} and 1047 {cm$^{-1}$}. 
  This doublet structure was not fully understood, although 
  the highest-energy component was assigned to the
  asymmetric proton-stretch fundamental based on the
  quantal calculations on the IR spectrum of {\mbox{H$_5$O$_2^+$}} of Bowman and coworkers.\cite{ham05:244301}
  Low-frequency modes may also play an important role in combination
  with the proton-transfer fundamental. 
  Such a possibility has been already suggested,\cite{fri04:9008,dai03:6571,sau:kal}
  but just which modes would participate in such combinations, and how, is still
  a matter of discussion.
 
\pagestyle{plain}

 The Hamiltonian used in the simulation of {\mbox{H$_5$O$_2^+$}} is expressed in a set of
 polyspherical coordinates based on the Jacobi vectors in Figure 1.\cite{gat99:7225}
 It is found that only after the introduction of such a curvilinear set of coordinates
 an adequate treatment of the
 anharmonic large-amplitude vibrations and torsions of the molecule becomes possible.
 The kinetic energy operator is {\em exact} for $J=0$, 
 and the derivation of its lengthy formula (674 terms) will be discussed 
 in a forthcoming publication. 
 The correctness of the operator implemented was checked by comparison
 with data generated by the TNUM program.\cite{lau02:8560}
 The internal coordinates used are: 
 the distance between the centers of mass of both water molecules ($R$), 
 the position of the central proton 
 with respect to the center of mass of the water dimer ($x$,$y$,$z$),
 the Euler angles defining the relative orientation between the two water 
 molecules (wagging or pyramidalization: $\gamma_a,\gamma_b$; rocking: $\beta_a,\beta_b$;
 internal relative rotation: $\alpha$) and the Jacobi coordinates
 which account for the particular configuration of each water 
 molecule ($r_{1(a,b)},r_{2(a,b)},\theta_{(a,b)})$) where $r_{1x}$ is the
 distance between the oxygen atom and the center of mass of the
 corresponding H$_2$ fragment, $r_{2x}$ is the H--H distance and
 $\theta_{x}$ is the angle between these two vectors. 
 These internal coordinates are body-fixed (BF) ones, where the water-water
 distance vector $\vec{R}$ points along the BF z-axis.
 These coordinates have the great advantage
 of leading to a much more decoupled representation of the
 PES than a normal-mode based Hamiltonian. 
 The quantum-dynamical problem is solved in the time-dependent picture using
 the multiconfiguration time dependent Hartree method 
 (MCTDH).\cite{man:bec,suppMCTDH}
 The potential energy surface has been represented by a cut high-dimensional 
 model-representation (cut-HDMR).\cite{bow:li,suppPES}

In Figure 2 probability-density projections on the wagging
coordinates are shown for the ground vibrational state ($g_0$), as well as
for one of the two fundamental states ($w_{1a}$,$w_{1b}$) of the wagging
modes, which are degenerate vibrational states with an energy of 106 {cm$^{-1}$}.
State $w_{3}$ is shown in
Figure 2c and it will play a major role due to its coupling to
the proton-transfer mode, as will be discussed later.

The probability-density of the wagging coordinates in $g_0$ (Figure 2a) 
presents four maxima in which the wagging angle is about 30 degrees with respect
to the planar conformation for each water molecule. The probability for one
or both water molecules to be found in a planar conformation is 
almost as high as the probability to be found pyramidal.
This means that {\mbox{H$_5$O$_2^+$}} interconverts already at $T=0$, due to zero-point energy,
between equivalent absolute minimum-energy structures in which both water molecules
are found in a pyramidal conformation.
Four equivalent minimum-energy structures are accessible through wagging motions.
The number of accessible equivalent minima at $T=0$ doubles to eight since the relative rotation
of both water molecules ($\alpha$ coordinate) has also been found to be allowed through a low energy
barrier. 

The energies of the next three wagging-mode states ($w_{2}$,$w_{3}$,$w_{4}$)
are, respectively, 232, 374 and 422 {cm$^{-1}$}.
In a harmonic limit these states can be represented by kets 
$|11\rangle$, \mbox{$(|20\rangle-|02\rangle)/\sqrt{2}$}
and \mbox{$(|20\rangle+|02\rangle)/\sqrt{2}$}, respectively, where the $|ab\rangle$ notation
signifies the quanta of excitation in the wagging motions of waters $a$ and $b$. 
The degeneracy between $w_{2}$, $w_{3}$ and $w_{4}$ is broken due to anharmonicity.
In harmonic approximation the energies
of the two lowest wagging-fundamentals $w_{1a}$ and $w_{1b}$ are about
300 {cm$^{-1}$} larger than our result and do not account for their
degeneracy, since harmonic modes are constructed taking as a reference only one of
the equivalent absolute minima. The system, however, interconverts between eight
equivalent $C_{2}$ structures and other stationary points through low-energy
barriers (wagging motions and internal rotation), which leads to a 
highly symmetric ground-state wavefunction.
Other vibrational states have been computed which are related to the
internal rotation, rockings and water-water stretching modes. They will
be reported and discussed in a forthcoming publication.

Figure 3 presents the IR predissociation spectrum
of the {\mbox{H$_5$O$_2^+$}}$\cdot$Ne complex \cite{ham05:244301} and the simulated spectrum
of {\mbox{H$_5$O$_2^+$}} in the range 700-1900 {cm$^{-1}$}.
The simulated spectrum is obtained in the time-dependent picture by Fourier 
transformation of the autocorrelation of a dipole-operated intial state:\cite{bal90:1741}
\begin{center}{\bf Eq. 1 HERE}\end{center}
where $E_0$ is the ground-state energy and
\mbox{$|\Psi_{\mu,0}\rangle \equiv \hat{\mu} \, |\Psi_0 \rangle$}.
The simulated spectrum shows a good agreement with the experimental spectrum.
The agreement on the doublet structure around 1000 {cm$^{-1}$} is very good,
and the position of the doublet at 1700 - 1800 {cm$^{-1}$} is also in good
agreement, despite the relative intensities being larger in the simulation.

The simulated spectrum in the range between 0 and 4000 {cm$^{-1}$} is depicted
in Figure 4. The region below 700 {cm$^{-1}$} has not yet
been accessed experimentally. Direct absorption of the wagging motions, 
excited by the perpendicular components of the field,
appears in the range between 100 - 200 {cm$^{-1}$}.
The doublet starting at 1700 {cm$^{-1}$} is clearly related to bending
motions of the water molecules, but its exact nature is still to be 
addressed. The simulated spectrum also
shows the absorptions of the OH stretchings starting at 3600 {cm$^{-1}$}.

 The doublet absorption at around 1000 {cm$^{-1}$} and the related underlying dynamics
 deserve a deeper analysis. 
 Due to the high density of states, it was not possible
 to obtain the fully
 converged states, but reasonably good approximations to the wavefunctions 
 of the low-energy \mbox{($|\Psi_d^l\rangle$, 930 {cm$^{-1}$})}
 and high energy \mbox{($|\Psi_d^h\rangle$, 1021 {cm$^{-1}$})} eigenstates of the doublet were computed.
 Even though these wavefunctions contain all the possible information on the two
 states, their direct analysis becomes complex due to the high dimensionality of such objects. 
 In order to obtain a fundamental understanding
 of the observed bands, zeroth-order states 
 $| \Phi_{z}\rangle$ and $| \Phi_{R,w_3}\rangle$
 were constructed, where $| \Phi_{z}\rangle$ is characterized by one quantum of excitation in the
 proton-transfer coordinate whereas $| \Phi_{R,w_3}\rangle$ by one quantum in
 the water-water stretch and two quanta in the wagging motion.
 They were constructed  
 by operating with $\hat{z}$ on the ground state: $|\Phi_{z}\rangle = \hat{z}|\Psi_0\rangle N$, 
 where $N$ is a normalization constant,
 and by operating with $(\hat{R}-R_0)$ on the third excited wagging state $w_3$:
 $|\Phi_{R,w_3}\rangle = (\hat{R}-R_0)|\Psi_{w_3}\rangle N$, respectively.
 The two eigenstates corresponding to the
 doublet were then projected onto these zeroth-order states.
 The corresponding overlaps read:
 $|\langle \Phi_{z} | \Psi_d^l \rangle|^2 = 0.20$, 
 $|\langle \Phi_{R,w_3} | \Psi_d^l \rangle|^2 = 0.53$ and
 $|\langle \Phi_{z} | \Psi_d^h \rangle|^2 = 0.48$,
 $|\langle \Phi_{R,w_3} | \Psi_d^h \rangle|^2 = 0.12$.
 One should take into account that these numbers depend on the exact
 definition of the zeroth-order states, which is not unique. 
 However, they provide a clear picture of the nature of the doublet:
 the low-energy band has the largest contribution from the combination of the
 water-water stretch and the third excited wagging (see Figure 2c), whereas
 the second largest is the proton-transfer motion. For the high-energy band
 the importance of these two contributions is reversed.
 Thus, the doublet may be regarded as a Fermi resonance between two zero-order
 states which are characterized by \mbox{($R$, $w_3$)} and \mbox{($z$)} excitations, respectively. 
 The reason why the third wagging excitation 
 plays an important role in the proton-transfer doublet is
 understood by inspecting Figure 2c and Figure 5.
 The probability density of state $w_3$ has four maxima, each of which
 corresponds to a planar conformation of \mbox{H$_2$O-H$^+$} (H$_3$O$^+$ character)
 for one of the waters, and a bend conformation (H$_2$O character) where
 a lone-pair H$_2$O orbital forms a hydrogen bond with the central proton. When the proton
 oscillates between the two waters, the two conformations exchange their
 characters accordingly.
Thus, the asymmetric wagging mode ($w_3$, 374 {cm$^{-1}$}) combines with
the water-water stretch motion ($R$, 550 {cm$^{-1}$}) to reach an energy close to the
natural absorption-frequency of the proton transfer. As a consequence, the
low-frequency wagging (or pyramidalization) motion of the water molecules
becomes strongly coupled to the higher frequency, spectroscopically active
proton-transfer motion and this coupling leads to the characteristic doublet feature of
the IR spectrum.

  In conclusion, we report a simulation of the dynamics and IR absorption
  spectrum of the {\mbox{H$_5$O$_2^+$}} cation by quantum-dynamical methodology
  in the full spectral range 0-4000 {cm$^{-1}$}. 
  The spectrum is directly comparable to available and future experiments on this system.
  We discuss some important features of the protonated water dimer which have remained
  until now elusive due to strongly anharmonic, large-amplitude motions.
  The floppy, multiminima nature of the cluster is presented and analyzed
  and the doublet-band absorption around 1000 {cm$^{-1}$} is fully reproduced
  and explained in terms of coupling of the proton-transfer motion to wagging torsions of the
  water moieties.
  These calculations constitute an avenue for a detailed quantum-dynamical description 
  of larger clusters 
  and provide important fundamental information on the spectroscopy
  and dynamics of protonated aqueous systems and the hydrated proton.


\clearpage

   \noindent {\bf Figure Captions:}\\[0.5cm]

   \noindent {\bf Figure 1}: Set of Jacobi vectors in terms of which the kinetic energy 
                             of the system is expressed. The set of internal 
                             coordinates used corresponds to the length of these 
                             vectors and relative angles. The $z$ direction of the
                             central proton is parallel to $\vec{R}$.

   \noindent {\bf Figure 2}: a) Probability density of the ground vibrational state, 
                             b) first and c) third wagging-mode states projected onto 
                             the wagging coordinates $\gamma_a$ and $\gamma_b$. 
            
   \noindent {\bf Figure 3}: a) Predissociation spectrum of the {\mbox{H$_5$O$_2^+$}}$\cdot$Ne complex;\cite{ham05:244301}
                             b) quantum-dynamical simulation. The resolution at which different peaks of the
                             spectrum are resolved is given by the Fourier Transform.
                             Due to the finite propagation time a finite resolution of about 30 cm$^{-1}$
                             is obtained.
                             
   \noindent {\bf Figure 4}: Quantum-dynamics simulated spectrum in the range between 0 and 
                             4000 {cm$^{-1}$}. Absorption is given in absolute
                             scale in mega-barns (Mb).

   \noindent {\bf Figure 5}: Two most important coupled motions
                             responsible for the doublet peak at 1000 {cm$^{-1}$}.

\clearpage
\pagestyle{empty}

\noindent {\bf Entry for the table of contents}\\[0.25cm]
          {\bf { Hydrated Proton}}\\[0.50cm]
       \begin{center}
       Oriol Vendrell, Fabien Gatti and Hans-Dieter Meyer*\\[0.25cm]
       {\bf Dynamics and Infrared Spectroscopy of the Protonated Water Dimer}\\[1cm]
       {\bf TOC GRAPHIC}\\[1cm]
       \end{center}
       {\bf The coupling shakes it:} dynamics and infrared absorption 
       spectrum of the protonated water dimer are reported by full quantum simulation.
       Strong couplings between the spectroscopically active proton-transfer motion and
       low-frequency, large-amplitude torsional modes are clearly identified and their 
       role in the cluster dynamics is explained.
       These couplings are responsible for the characteristic doublet-peak around 1000 cm-1, 
       which was not understood and subject of debate. This spectral feature is
       reproduced, assigned and explained.

       \vspace{1.8cm}
       \noindent {\bf keywords:} water clusters $\cdot$ proton transport $\cdot$ 
                                 IR spectroscopy $\cdot$ quantum dynamics

 \clearpage
 \begin{center}
 \includegraphics[scale=1.0]{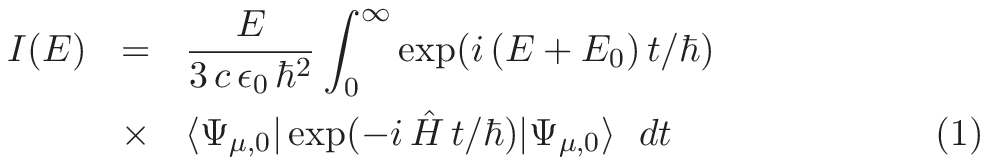}
 \end{center}

 \clearpage
 \begin{center}
 \includegraphics[scale=1.0]{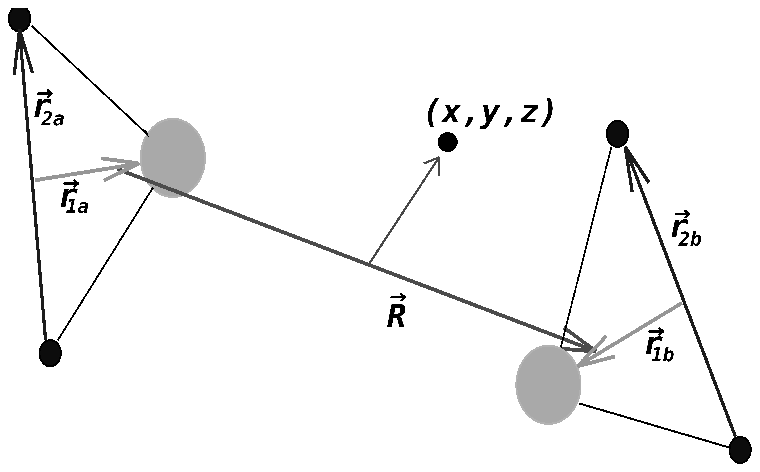}
 \end{center}

 \clearpage
 \begin{center}
 \includegraphics[scale=1.0]{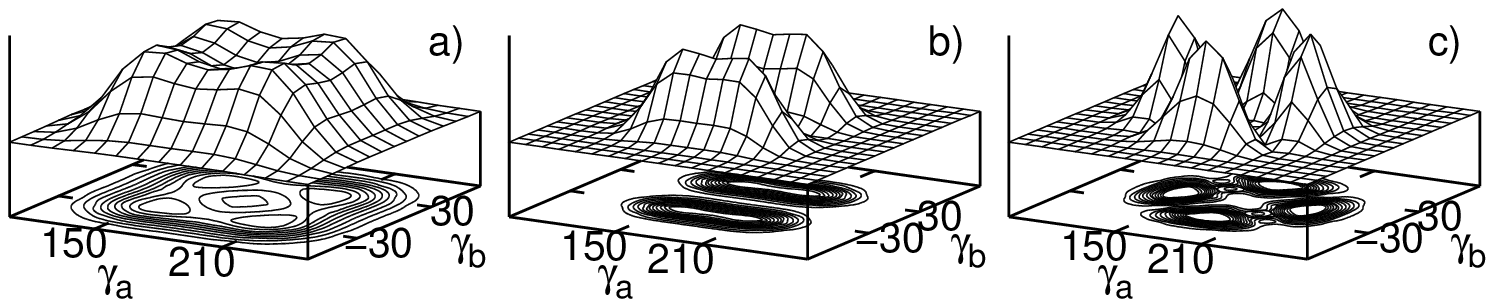}
 \end{center}

 \clearpage
 \begin{center}
 \includegraphics[scale=1.0]{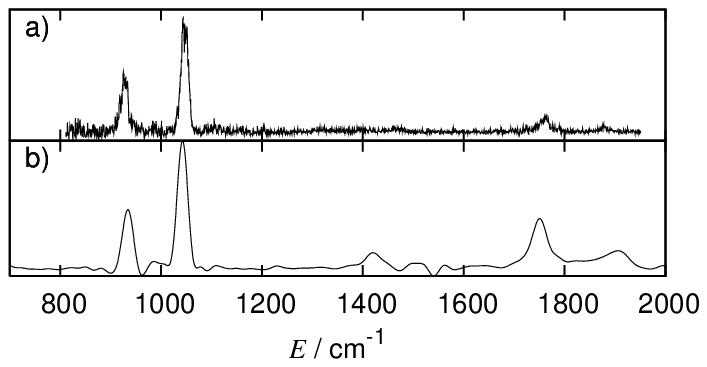}
 \end{center}
 
 \clearpage
 \begin{center}
 \includegraphics[scale=1.0]{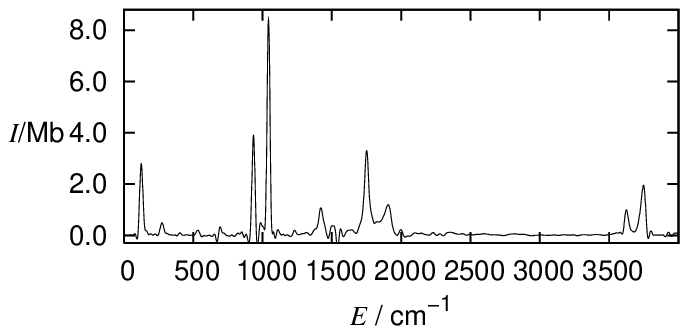}
 \end{center}

 \clearpage
 \begin{center}
 \includegraphics[scale=1.0]{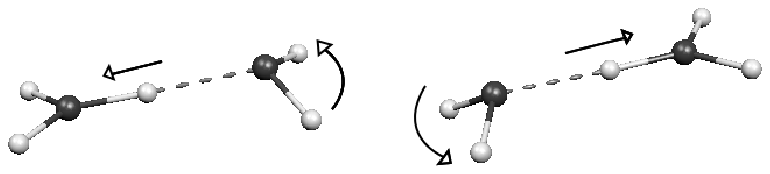}
 \end{center}

 \clearpage
 \begin{center}
 \includegraphics[scale=1.0]{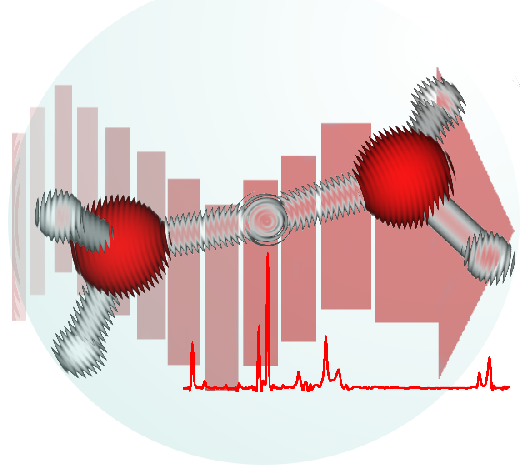}
 \end{center}

\end{document}